\documentclass[11pt]{article}
\usepackage{moriond}

\def\be{\begin{equation}}
\def\ee{\end{equation}}
\def\bea{\begin{eqnarray}}
\def\eea{\end{eqnarray}}

\newcommand{\like}{\mathcal{L}}
\newcommand{\pdf}{\mathcal{P}}
\newcommand{\Emin}{{E_{\mathrm{min}}}}
\newcommand{\Emax}{{E_{\mathrm{max}}}}
\newcommand{\Photo}{}

\begin{document}
\hfill IFIC/15-28, IPPP/15/25, DCPT/15/50
\vspace*{4cm}
\title{The Flavour Composition of the High-Energy IceCube Neutrinos}

\author{ Aaron C. Vincent}
\address{Institute for Particle Physics Phenomenology (IPPP),\\ Department of Physics, Durham University, Durham DH1 3LE, UK.}
\author{Sergio Palomares Ruiz, Olga Mena}
\address{Instituto de F\'{\i}sica Corpuscular (IFIC)$,$
 CSIC-Universitat de Val\`encia$,$ \\  
 Apartado de Correos 22085$,$ E-46071 Valencia$,$ Spain}
\maketitle\abstracts{
We present an in-depth analysis of the flavour and spectral composition of the 36 high-energy neutrino events observed after three years of observation by the IceCube neutrino telescope. While known astrophysical sources of HE neutrinos are expected to produce a nearly $(1:1:1)$ flavour ratio (electron : muon : tau) of neutrinos at earth, we show that the best fits based on the events detected above $E_\nu \ge 28$ TeV do not necessarily support this hypothesis. Crucially, the energy range that is considered when analysing the HE neutrino data can have a profound impact on the conclusions. We highlight two intriguing puzzles: an apparent deficit of muon neutrinos, seen via a deficit of track-like events; and an absence of $\bar \nu_e$'s at high energy, seen as an absence of events near the Glashow resonance. We discuss possible explanations, including the misidentification of tracks as showers, and a broken power law, in analogy to the observed HE cosmic ray spectrum.}

\section{Introduction}
The detection of 37 high-energy events consistent with an astrophysical origin at the IceCube neutrino detector at the South Pole\cite{Aartsen:2013bka,Aartsen:2013jdh,Aartsen:2014gkd} has signalled the beginning of the era of high-energy neutrino astronomy. Neutrinos are a powerful, complementary tool in observations of the extragalactic sky: unlike protons and other cosmic rays, they are not affected by magnetic fields in the intergalactic medium and thus point directly back towards their source; their low interaction cross-section means that they are not attenuated during their propagation to earth. In principle they allow us to peer \textit{inside} sources. Finally, they carry an additional quantum number, \textit{flavour}, which can yield information about source properties, propagation and detection, and has long been known to be an important observable for both astro and particle physics\cite{Athar:2000yw,Beacom:2003nh,Kashti:2005qa,Rodejohann:2006qq,Lipari:2007su,Pakvasa:2007dc}.

The flavour composition is determined by the production mechanism and by oscillation during propagation. The canonical astrophysical neutrino production scenario is via the disintegration of charged pions produced by proton-proton and photon-proton collisions in cosmic ray sources. This yields a flavour composition at the source $(\alpha_e : \alpha_\mu : \alpha_\tau) = (1:2:0)$. Other scenarios can affect this expectation: rapid muon energy loss can suppress the production of $\nu_e$'s at the same energy scale, giving $(0:1:0)$, while the the same phenomenon at higher energies gives the complementary ratio $(1:1:0)$. Finally, neutron decay-dominated sources yield only electron (anti) neutrinos, giving $(1:0:0)$. Following production, neutrino oscillation averages the flavour composition over the large uncorrelated distances to Earth. The $(1:2:0)$ canonical expectation averages to a ratio very close to $(1:1:1)$ at Earth; other source compositions must lie inside a small triangle around this point (shown in blue in our figures below). A measured flavour composition at Earth outside of this triangle would indicate either of exotic new physics during propagation,  a non-astrophysical origin, or a systematic problem in the flavour reconstruction, e.g. due to a misidentification of event topologies. 

The search for astrophysical neutrinos is complicated by the large production rate of atmospheric neutrinos, originating from the pions and kaons produced when cosmic rays strike the Earth's atmosphere. These are detected in IceCube at a rate of $\sim$ 3000 per second, dwarfing the expected astrophysical signal by orders of magnitude. Atmospheric neutrinos are characterized by a high coincidence rate with muons and other neutrinos, anisotropies due to the projected atmospheric column density, as well as characteristic flavour and spectral information. By establishing a strict veto protocol, the IceCube collaboration was able to isolate 37 events in 3 years of data, consistent with neutrinos from a non-atmospheric origin, with energies between 28 TeV and 2 PeV. These events follow the expected shallower spectral shape, and are so far consistent with an isotropic distribution in the sky.

In this study \cite{Mena:2014sja,Palomares-Ruiz:2014zra,Palomares-Ruiz:2015mka}, we focused on the flavour composition of the observed high-energy neutrinos. Our first study has spurred wide interest both from the IceCube collaboration \cite{Aartsen:2015ivb,JuanPablo} as well as independent authors \cite{Anchordoqui:2014pca,Watanabe:2014qua,Palladino:2015zua}. TeV--PeV energy neutrinos are seen in IceCube via one of two morphologies: 1) showers (cascades), caused by neutral current (NC) interactions with nuclei, and charged current (CC) interactions of electron and tau neutrinos in the ice; and 2) muon tracks, which trace the propagation of a high-energy muon produced by the CC interaction of a muon neutrino, or by tau lepton production followed by decay to muon. Out of three years of data collection, IceCube expects 8.4 track-producing atmospheric muons and 6.6 atmospheric neutrinos. However, only 8 out of the 36 \footnote{We discard event 32 because its energy could not be reconstructed and was coincident with a pair of background atmospheric muons.} observed events are muon tracks. From these numbers alone, there is a hint of a possible deficit in astrophysical muon neutrinos.

In this work, we use both the topological and spectral information available for the 36 high-energy neutrinos seen at IceCube to test the available parameter space of flavours, spectra and total contribution of astrophysical and atmospheric neutrinos and muons to observations.

We begin with an overview of the high energy neutrino event rate calculation is Sec.~\ref{sec:theory}, followed in Section \ref{sec:results} by our likelihood analysis.  We conclude with a short discussion of the physical implications and of other flavour studies that have appeared in the recent literature. 

\section{Spectral analysis of the IceCube events}
\label{sec:theory}

The spectrum of true deposited energies ${dN^c}/{dE_{\mathrm{true}}}$  -- the electromagnetic energy that is deposited in the ice -- as a function of incident neutrinos of energy $E_\nu$, flavour $\ell = \{e,\mu,\tau\}$ and origin $f$ (astrophysical or atmospheric), is  a function of the exposure $T$, the whole-sky averaged attenuation $Att^f_{\nu_\ell}(E_\nu)$, the effective mass $ M_{\mathrm{eff}}(E_{\mathrm{true}})$ of the detector and the differential neutrino flux ${d\phi^f_{\nu_\ell}(E_\nu)}/{d E_\nu}$. It follows a general form:
\begin{equation}
\frac{dN^c}{dE_{\mathrm{true}}} = T \, N_A \, \int_0^\infty Att^f_{\nu_\ell}(E_\nu) \, M_{\mathrm{eff}}(E_{\mathrm{true}}) \, \frac{d\phi^f_{\nu_\ell}(E_\nu)}{d E_\nu} \, \frac{d\sigma^c_{\nu_\ell}(E_\nu,E_{\mathrm{true}})}{dE_{\mathrm{true}}} \, d E_\nu ~. 
\label{eq:dNdEd}
\end{equation}
The cross section ${d\sigma^c_{\nu_\ell}(E_\nu,E_{\mathrm{true}})}/{dE_{\mathrm{true}}}$ depends on the neutrino flavour and the interaction channel $c$. If the cascade is electronic, all of the energy is effectively deposited in the detector; if it is hadronic, then the energy-dependent deposition efficiency must be included in the calculation of (\ref{eq:dNdEd}). The specific form of the cross sections and differential spectra in (\ref{eq:dNdEd}) is given in the appendices of our detailed study~\cite{Palomares-Ruiz:2015mka}. In the case of CC muon neutrino events, the energy deposited by the muon track must be added to the associated cascade. We model this by computing the average energy deposited by a muon with a random point of origin and orientation in the detector. At relevant energies, this is equal to 0.119 times its initial energy. We account for the escape of tau leptons from the fiducial volume before decay in a similar manner. 

We also include interactions between antielectron neutrinos and electrons in the ice, although this only becomes important around the Glashow resonance at 6.3 PeV, where the centre of mass energy is equal to the W boson mass. 

The effective detector mass as a function of $E_{\mathrm{true}}$ is obtained by deconvolving the effective masses provided by IceCube\cite{Aartsen:2013bka} as a function of the neutrino energy $E_{\nu}$. The flux in terms of deposited energy in the detector is then evaluated as a function of the ``true'' electromagnetic energy deposited in the ice via a convolution with a resolution function $R(E_{true},E_{dep}, \sigma(E_{true}))$:
\begin{equation}
\frac{dN^c}{dE_{\mathrm{dep}, i}} = \int_0^\infty \frac{dN^c}{dE_{\mathrm{true}}} \, R(E_{\mathrm{true}},E_{\mathrm{dep},i},\sigma(E_{\mathrm{true}})) \, d E_{\mathrm{true}} ~.
\label{eq:dNdEde}
\end{equation}
We model $R$ as a gaussian distribution around $E_{\mathrm{dep}}$ with an asymmetric standard deviation, fitted to the errors given for each event \cite{Aartsen:2014gkd}.

Atmospheric differential event rates are modelled in the same way, using recent computations of the atmospheric fluxes\cite{Sinegovskaya:2014pia}. These are then averaged over the whole sky, and a suppression due to IceCube's veto is included. Attenuation of all flavours and regeneration of tau neutrinos in the passage through Earth is included for astrophysical and atmospheric events.

\section{Results}
\label{sec:results}
We turn to our main results. The reader interested in more detailed results -- including tabulated best fits and exclusions, as well as results concerning a prompt (charm) atmospheric component, is referred to the complete paper \cite{Palomares-Ruiz:2015mka}.

\begin{figure}[h]
%\begin{tabular}{ c c}
\includegraphics[width=.5\textwidth]{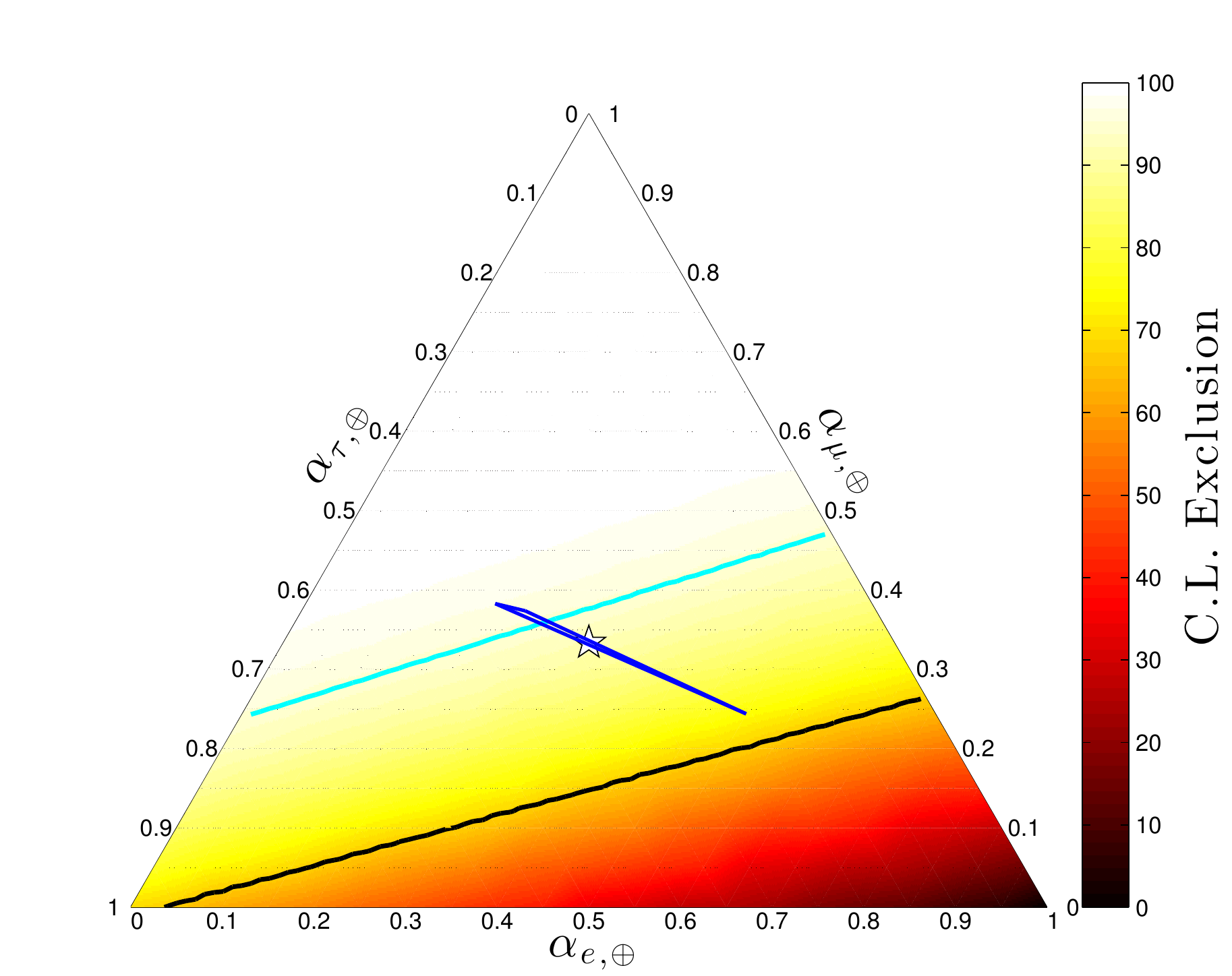}  \includegraphics[width=.5\textwidth]{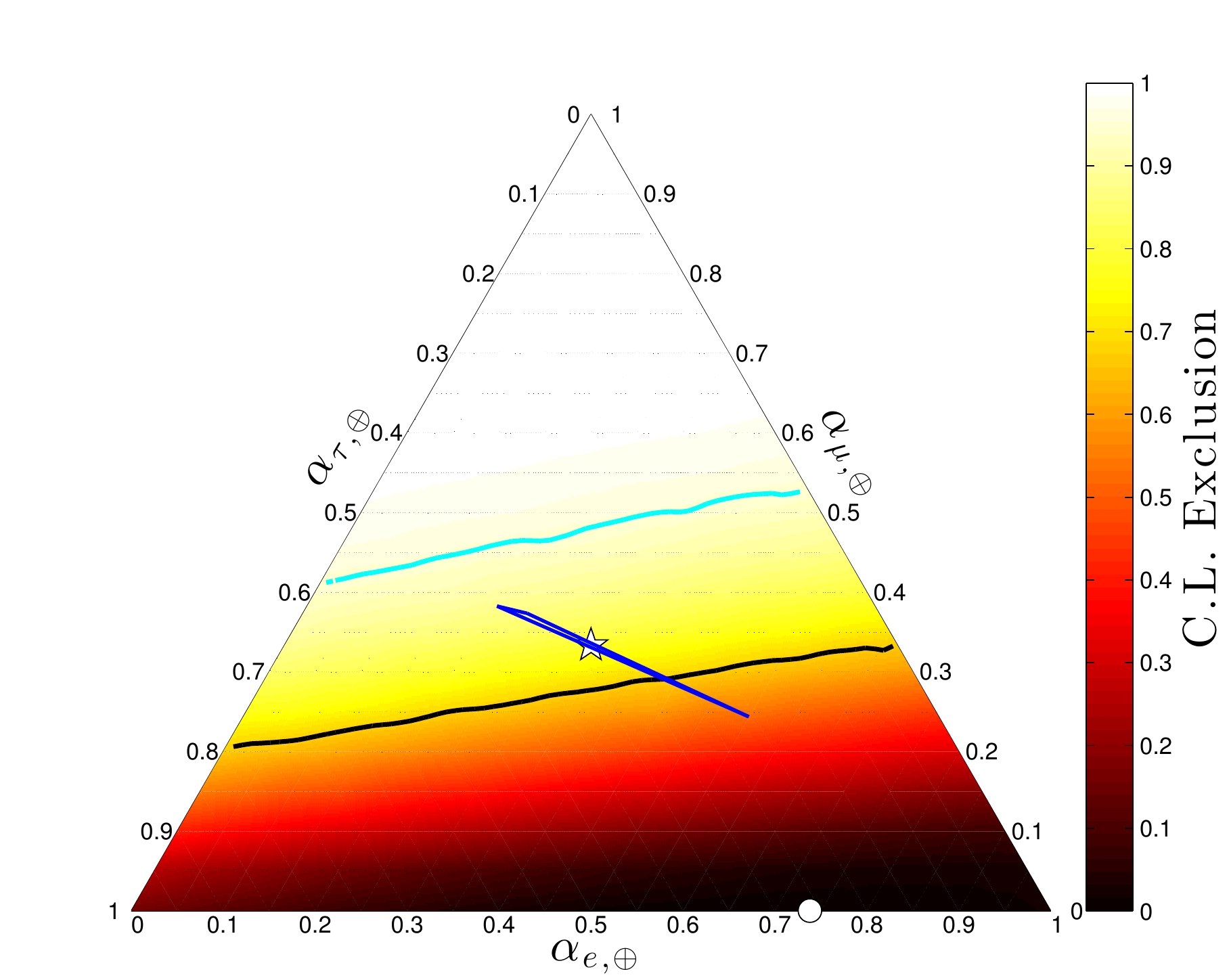} 
%\end{tabular}
\caption{Left: the exclusion confidence limits on possible flavour compositions for the 36 high-energy events observed at IceCube in the 28 TeV - 2 PeV range, using event topology information alone, fixing the atmospheric muon and neutrino rates to their expected values, and fixing the spectral index to $\gamma = 2$. The flavour composition and number of astrophysical neutrinos $N_a$ is allowed to vary. Right: same as left panel, but including spectral information as in (\ref{eq:pdf}-\ref{eq:like}). In every case the thin blue triangle indicates the space to which astrophysical neutrinos may oscillate, the white star indicates $(1:1:1)$ and the best fit is denoted by a white circle.}
\label{fig:1}
\end{figure}

We start by restricting ourselves to a study of the topological information in the 36 observed events \cite{Mena:2014sja,Palomares-Ruiz:2014zra}. The likelihood $\like$ of the observed event topology composition as a function of the flavour composition, is presented in the left panel of Fig. \ref{fig:1}. We have fixed the background atmospheric muon and neutrino fluxes to the expected values (8.4 and 6.6, respectively), and the astrophysical neutrino spectrum to $\phi_\nu \propto E_\nu^{-2}$. The black line is the 68\% CL exclusion line; cyan represents 95\% CL. The best fit occurs at $(1:0:0)$, and the canonical $(1:1:1)$ is disfavoured at 92\% CL. Assuming a slightly steeper spectrum somewhat softens this constraint. This strong constraint is due to the low track-to-shower ratio $(8 : 28)$ which is taken up entirely by the expected atmospheric muon background. 

In order to include spectral information, we introduce the expected flux (\ref{eq:dNdEd}) into the following PDF, as a function of the flavour composition $\{\alpha_\ell\}$, for each event $i$ of topology $k$ and caused by each type $f$ of incoming particle (astrophysical, atmospheric muon or atmospheric neutrino):
\begin{equation}
\pdf_{i}^{ f} (\{\alpha\}, \gamma) = \frac{1}{\sum_{\ell, j} \alpha_\ell \int_\Emin^\Emax dE_{\mathrm{dep}} \,  \frac{dN_{\ell}^{j, f}}{dE_{\mathrm{dep}} }} \, \sum_\ell \alpha_\ell \, \frac{dN_{\ell}^{ f}}{dE_{\mathrm{dep},i}}   ~.
\label{eq:pdf}
\end{equation}
From the partial likelihood $\like_i = \sum_f N_f \pdf_{i}^{f}$, the total likelihood is then:
\begin{equation}
\like = e^{-N_a  - N_\nu - N_\mu} \prod_{i=1}^{N_{\mathrm{obs}}}  \like_i ~.
\label{eq:like}
\end{equation}
The right panel of Fig.~\ref{fig:1} shows the effect of including spectral information. The best fit is still close to $(1:0:0)$, but constraints are weaker: this is an indication that while the low number of tracks is entirely consistent with a complete atmospheric origin, their spectrum is not. 

\begin{figure}[h]
%\begin{tabular}{ c c}
\includegraphics[width=.5\textwidth]{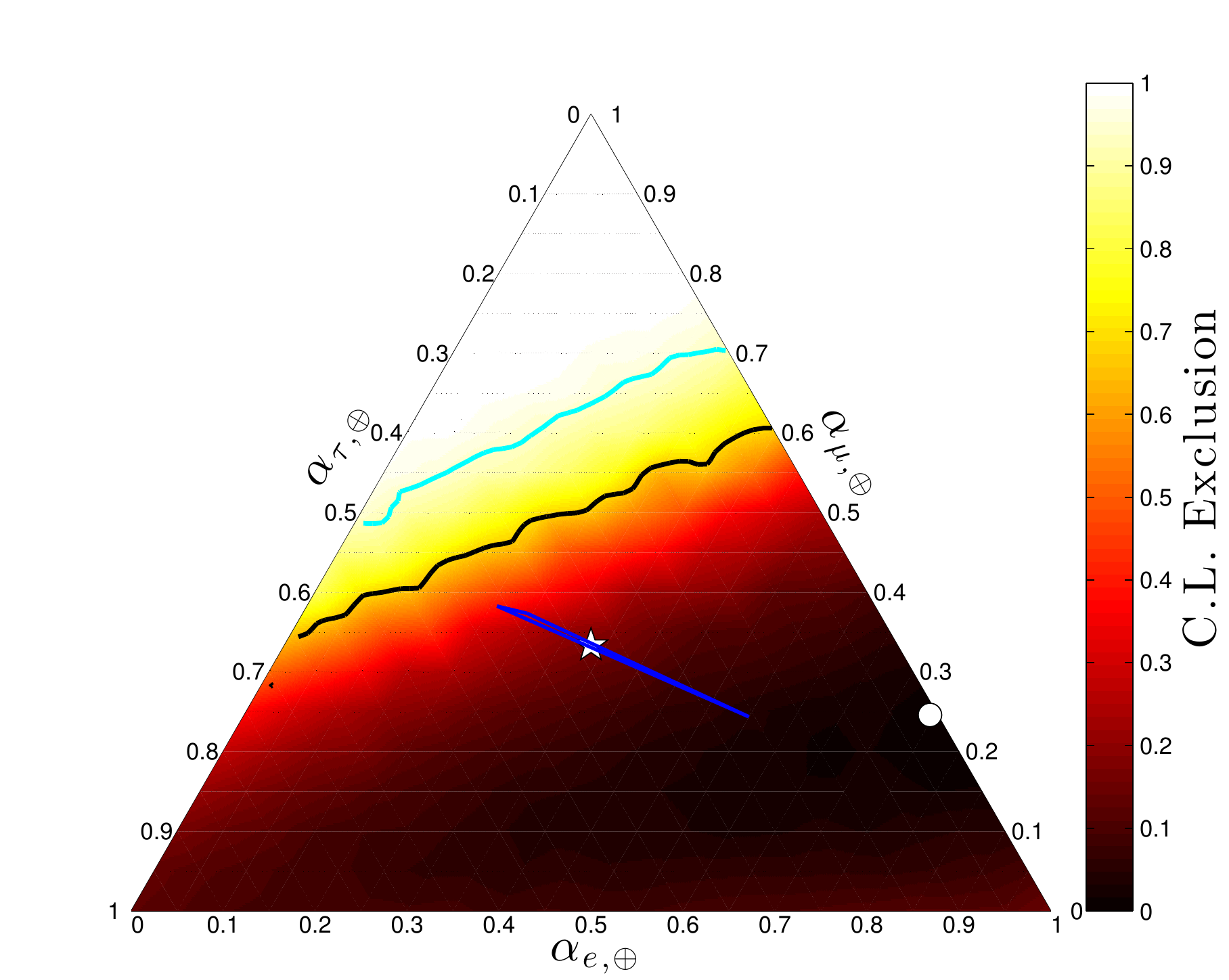}  \includegraphics[width=.5\textwidth]{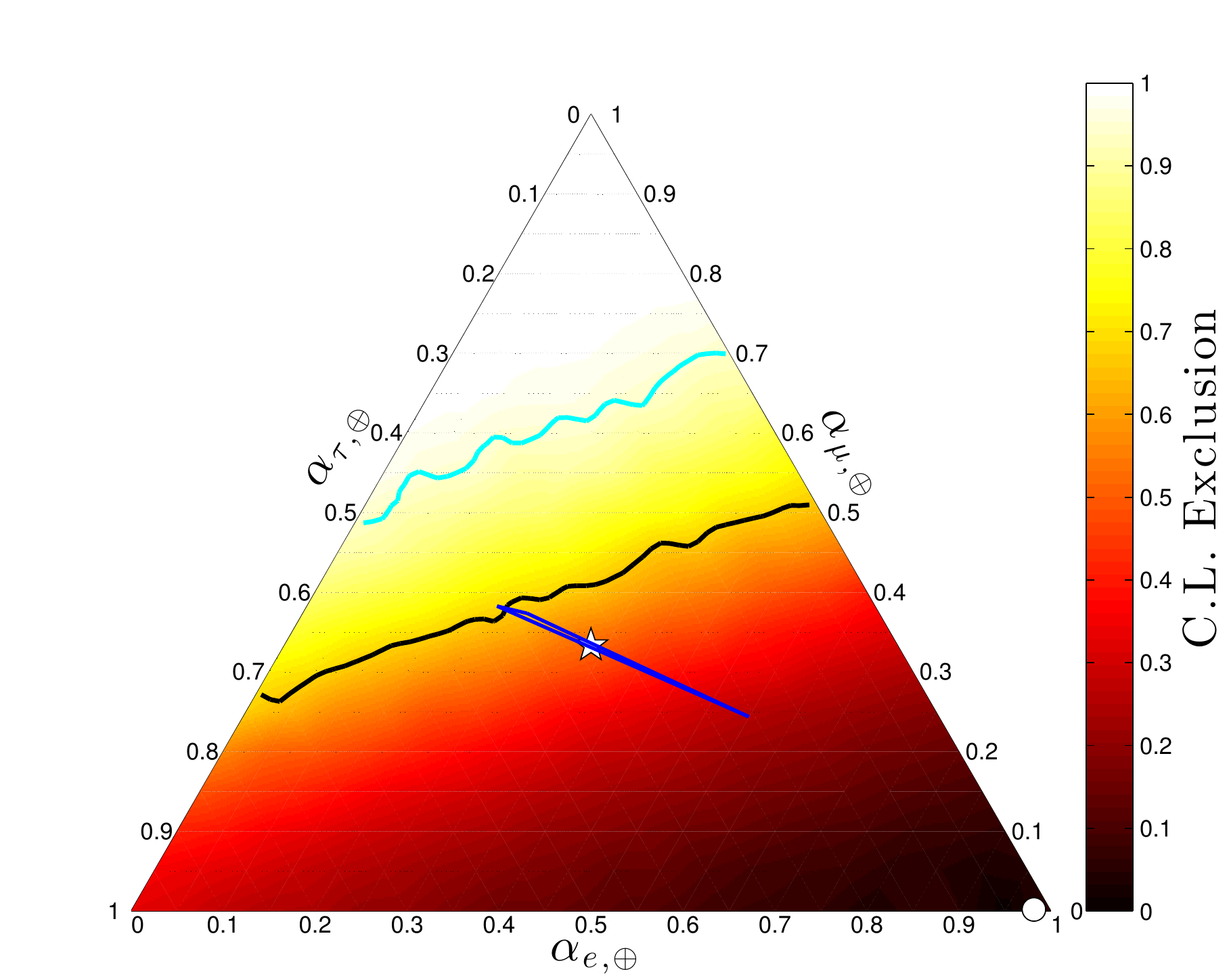} 
%\end{tabular}
\caption{As in Fig. \ref{fig:1}, but allowing the atmospheric muon $N_\mu$ and neutrino fluxes $N_\nu$, the astrophysical spectral index $\gamma$ to vary, in addition to the flavour composition and astrophysical flux $N_a$. Right panel: same, but only considering the 20 events observed above 60 TeV. }
\label{fig:2}	
\end{figure}

This can be tested by freeing the atmospheric flux $N_\mu$ and $N_\nu$. We also allow the astrophysical flux's spectral index $\gamma$ to vary freely. Results are shown in the left panel of Fig.~\ref{fig:2}. In this case a wider range of flavour ratios may be accommodated: the spectral index is best fit by $\gamma = 2.96$, and $N_\mu = 4.7$, $N_\nu = 4.8$. If the energy range is cut at $\Emin = 60$ TeV, most of the atmospheric background is removed, leaving only 20 events and a cleaner astrophysical signal. Interestingly, in this case the best fit returns to $(1:0:0)$. This is shown in the right panel of Fig.~\ref{fig:2}. One recovers $\gamma = 2.34$, very near IceCube's best fit ($\gamma = 2.3$), and atmospheric neutrinos should account for 6.5 events, while the best fit for atmospheric muons is consistent with expectations, with $N_\mu = 0.1$. 

\begin{figure}[h]
%\begin{tabular}{ c c}
\includegraphics[width=.5\textwidth]{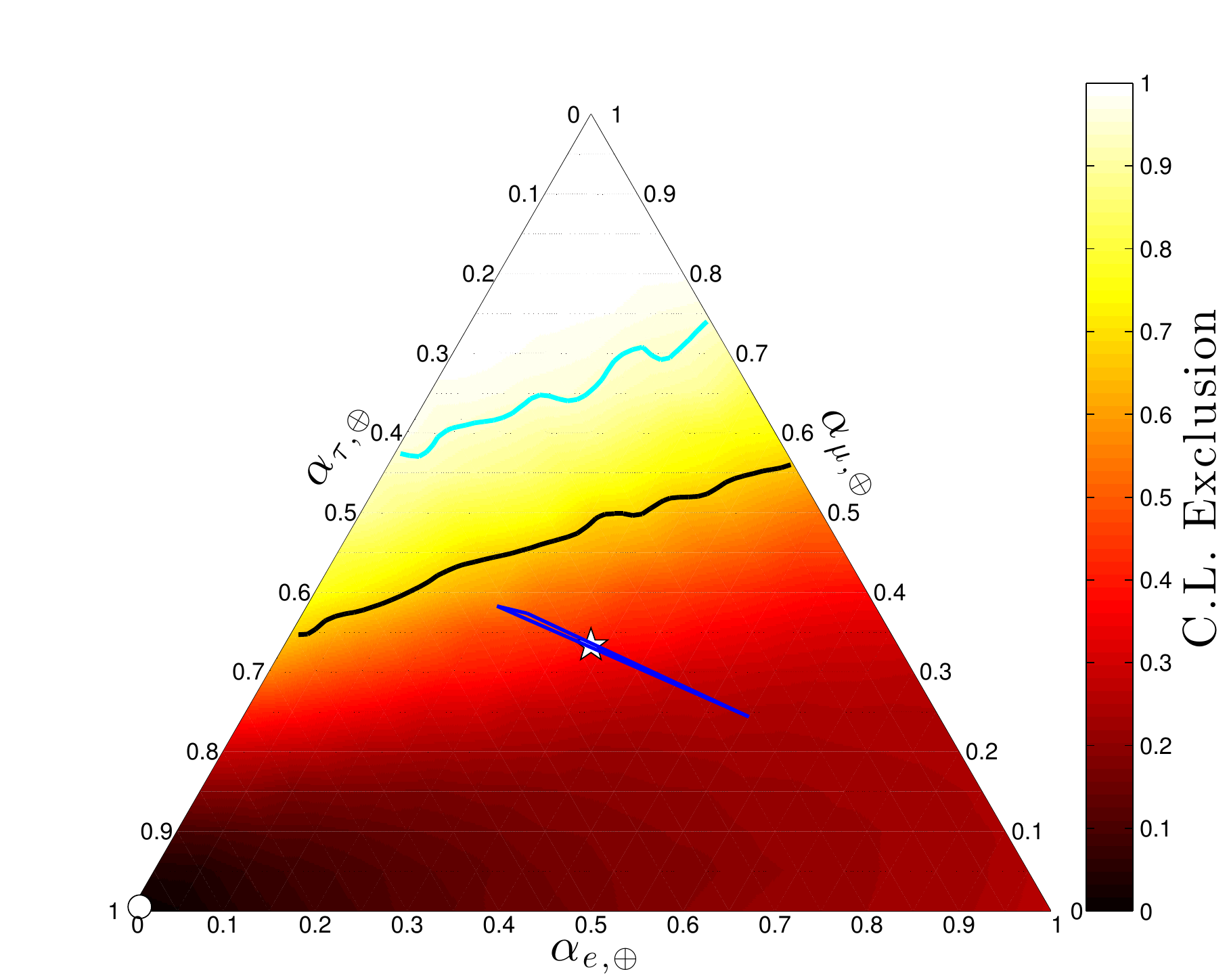}  \includegraphics[width=.5\textwidth]{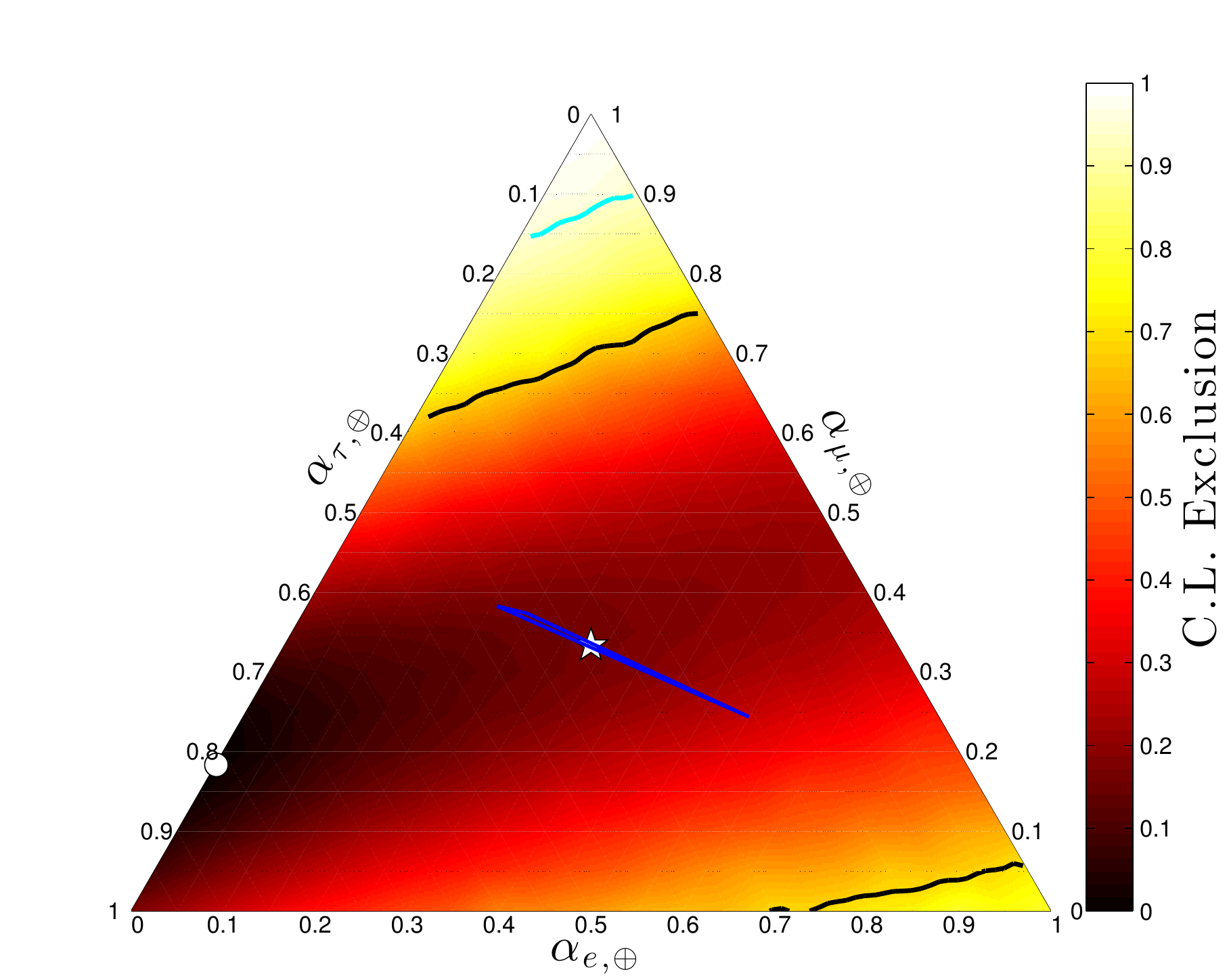} 
%\end{tabular}
\caption{As in Fig. \ref{fig:2}, but extending the energy range to 60 TeV -- 10 PeV. Left: the absence of events near the Glashow resonance leads to a best fit that is mostly composed of tau neutrinos. Right panel: the effect of assuming a 30\% rate of misidentification of muon tracks as showers.}
\label{fig:3}	
\end{figure}

There is no a priori reason to cut the energy range at $\Emax = 2$ PeV. In fact, at 6.3 PeV a larger flux is expected from the Glashow resonance. We show the effect of increasing the energy range to 10 PeV in the left panel of Fig.~\ref{fig:3}. The best fit in the 60 TeV -- 10 PeV range is near $(0:0:1)$: an indication of a deficit in muon neutrinos (from the topological information) and in electron antineutrinos (from the lack of events at the Glashow resonance). In this case $\gamma = 2.48$, $N_\nu = 1.5$ and $N_\mu = 2.2$. 


\section{Discussion}
From the results presented in Figs.~\ref{fig:1}--\ref{fig:3}, we are left with two potential puzzles, assuming future IceCube events follow the trends seen here. The first is a paucity of muon tracks, indicating a lack of muon neutrinos in the astrophysical flux. The second is a deficit of electron antineutrinos around the Glashow resonance. The latter can be solved for example by a break in the power law, in analogy with the cosmic ray spectrum. The former can lead to more interesting effects. If the best fit ends up away from $(1:1:1)$, it could mean that a non-standard source is responsible for the bulk of the astrophysical neutrino flux, e.g. neutron decay\cite{Anchordoqui:2014pca}. If the best fit lies outside the thin blue region, indicating a flavour composition that cannot be achieved by known oscillation physics, this could be an indication a non-astrophysical origin, or of new phenomena such as neutrino decay\cite{Beacom:2002vi}, extra dimensions\cite{Aeikens:2014yga}, modifications of gravity\cite{Illana:2014bda} or other new effects during propagation. The explanation could be much more mundane, however. A misidentification of muon tracks as showers could easily account for the $\nu_\mu$ deficit\cite{Mena:2014sja,Palomares-Ruiz:2015mka,Aartsen:2015ivb}. We show the effect of a 30\% misidentification rate in the right panel of Fig. \ref{fig:3}. In this case, compatibility with $(1:1:1)$ is much more plausible.
	
We have shown that the assumptions that go into the reconstruction of the flavour and spectral composition of the 36 high energy neutrino events observed at IceCube are crucial when it comes to drawing conclusions about their origin. Other studies of the flavour composition ostensibly agree with the results presented here\cite{Palomares-Ruiz:2015mka}. A recent study by IceCube\cite{Aartsen:2015ivb}  agreed with our early conclusions after repeating our single-energy bin analysis\cite{Mena:2014sja,Palomares-Ruiz:2014zra}. After adding 101 lower-energy events, they found a preference for a tau-dominated flux when including spectral information and extending the analysis to 10 PeV. They attribute the lack of $\nu_\mu$'s to a 30\% misclassification rate of tracks as showers, in agreement with our findings in Fig.~\ref{fig:3}, and with our initial conclusions\cite{Mena:2014sja}. In a separate study, the inclusion of a larger number of low-energy events consistent with an astrophysical component gives a similar picture, but flipped so that the best fit lies along the $\nu_e-\nu_\mu$ axis, rather than the $\nu_\tau-\nu_\mu$ axis\cite{JuanPablo}. This is unsurprising, since adding several hundred events below 100 TeV removes any statistical power from the Glashow resonance region.

Nonetheless, the significance of the results presented here and in other analyses remain low -- even when through-going muons are included\cite{Palladino:2015zua} -- for the simple reason of low statistics. While IceCube expects a few dozen more events in the next years, future experiments such as KM3Net and Gen-2 IceCube will be crucial in the next step of the new exciting field of neutrino astronomy.

\section*{Acknowledgments}
 S.~P.~R. is supported by a Ram\'on y Cajal contract, by the Spanish MINECO under Grant No. FPA2011-23596, by the Generalitat Valenciana under Grant PROMETEOII/2014/049, and  by the Portuguese FCT through CERN/FP/123580/2011 and PTDC/FIS-NUC/0548/2012, partially funded through POCTI (FEDER). O.~M. is supported by the Consolider Ingenio Project CSD2007--00060 and by Grant FPA2011--29678 of the Spanish MINECO, and by PROMETEO/2009/116. The authors are also partially supported by PITN-GA-2011-289442-INVISIBLES.
%\section*{Appendix}
%
% We can insert an appendix here and place equations so that they are
%given numbers such as Eq.~\ref{eq:app}.
%\be
%x = y.
%\label{eq:app}
%\ee

\section*{References}

\bibliography{flavors}

\end{document}